\documentstyle[preprint,aps]{revtex}

\newcommand{\bra}[1]{\left\langle #1 \right|}
\newcommand{\ket}[1]{\left| #1 \right\rangle}

\begin{document}

\draft
\title{Asymmetric nuclear matter and neutron star properties}
\author{L.\ Engvik, M.\ Hjorth-Jensen and E.\ Osnes}
\address{Department of Physics,
University of Oslo, N-0316 Oslo, Norway}
\author{G.\ Bao and E. \O stgaard}
\address{Department of Physics, AVH,
University of Trondheim, N-7055 Dragvoll, Norway}

\maketitle

\begin{abstract}
We calculate properties of neutron stars such as mass
and radius using a relativistic Dirac-Brueckner-Hartree-Fock
approach. Modern meson-exchange potential models are used to
evaluate the $G$-matrix for asymmetric nuclear matter.
For pure neutron matter we find the maximum mass to
be $M_{\mathrm{max}}\approx 2.4 M_{\odot}$ for a radius $R\approx 12$ km,
whereas with a proton fraction of $30\%$, we find
$M_{\mathrm{max}}\approx 2.1 M_{\odot}$ for a radius $R\approx 10.5$ km, close
to the experimental values.  The implications
are discussed.
\end{abstract}


\clearpage

The properties of neutron stars depend on the equation of state (EOS)
at densities up to an order of magnitude higher than those observed
in ordinary nuclei.
Data on the EOS can be obtained from many sources, such as studies of
the monopole resonance in finite nuclei, high-energy
nuclear collisions, supernovae and neutron  stars.
Supernova simulations seem to require  an EOS which is too soft to
support some observed masses of neutron stars, whereas analyses of
high-energy nuclear collisions indicate a rather stiff EOS, predicting
neutron star masses which are too large.
Thus, no definite statements can be made
about the EOS at high densities, except that it
should probably be moderately stiff in order to
support maximum neutron star masses in a range
from approximately $1.4 M_{\odot}$ to $1.9 M_{\odot}$ \cite{thorsett93}.

However, although quantitative calculations of the EOS for dense
nuclear matter are still beset with many problems, there have recently
been important changes in the qualitative picture. Several mechanisms
have been studied, mechanisms which
result in an EOS which is soft enough to support
neutron stars with maximum masses in the range of the observed data.
Among such processes
we find exotic states of nuclear matter, such as
kaon
\cite{kn87,brown93} or pion condensation \cite{mo88}.
Another scenario which gives neutron star masses
within the experimental
values, has been presented by Pethick and co-workers
\cite{lrp93,hps93}. These authors study
properties of various phase transitions
from spherical nuclei to uniform nuclear matter, and the coexistence
of quark matter and nuclear matter over a finite fraction of the
neutron star volume.

The scope of this work is to derive the EOS for asymmetric nuclear
matter, using relativistic many-body theories within the
framework of the Dirac-Brueckner-Hartree-Fock approach
\cite{cs86,bm90} and treating the Pauli operator which enters
our formalism (see below) correctly. To our knowledge, this has
not been done before.
Asymmetric nuclear matter is important in e.g.\ studies of neutron
star cooling, as demonstrated recently by Lattimer {\em et al.}
\cite{lpph91} who showed that ordinary nuclear matter
with a small asymmetry parameter can cool by the so-called
direct URCA process even more rapidly than matter in an exotic state.
In addition, at high densities, degrees of freedom represented by
isobars and hyperons may also result in a rapid cooling of neutron stars
\cite{pplp92}.

Before we present our results, we briefly sketch below our calculational
procedure, for further details see Ref.\ \cite{behoo94}.
Important
ingredients in our calculations are the nucleon-nucleon (NN)
interaction which we take from meson-exchange models, and the
renormalized NN potential in a medium like nuclear matter.
The latter is accounted for by  the reaction
matrix $G$
\begin{equation}
   G(\omega )=V+VQ\frac{1}{\omega - QH_0Q}QG(\omega ),
   \label{eq:bg}
\end{equation}
where $\omega$ is the energy of the interacting  nucleons,
$V$ is the free NN potential, $H_0$ is the unperturbed energy of the
intermediate scattering states,
and $Q$ is the Pauli
operator which prevents scattering into occupied states.
The $G$-matrix is used in our many-body scheme described
below in order to obtain the energy per particle in asymmetric
nuclear matter.
To calculate the NN potential, we use the Bonn A potential defined
in table A.1 of Ref.\ \cite{mac89}. This potential was shown by Brockmann
and Machleidt \cite{bm90},
to reproduce the nuclear matter binding energy and saturation density
within the framework of the Dirac-Brueckner-Hartree-Fock approach.

In order to test whether our EOS is appropriate
for neutron stars, we evaluate the total mass and radius of such
stars. Although there are recent relativistic calculations which
address asymmetric nuclear matter,
see e.g.\ the recent work of Huber {\em et al.}
\cite{hww93}, one of the  differences
between this work and Ref.\ \cite{hww93}
is an exact treatment of the Pauli operator $Q$ for asymmetric
nuclear matter. The conventional approach has been to extrapolate
the results for symmetric nuclear matter and pure neutron matter
\cite{bi92}. Here we employ an exact expression for $Q$
in nuclear matter as described in e.g.\ Ref.\ \cite{swk92}
and calculate the energy per particle for asymmetric nuclear
matter using an extended Brueckner-Hartree-Fock (BHF) method,
see Ref.\ \cite{ms89} for a discussion of the non-relativistic
BHF theory,
namely the Dirac-Brueckner-Hartree-Fock (DBHF) method.
The DBHF method is a variational procedure, where the single-particle (sp)
energies are obtained through an iterative self-consistency scheme.
To describe the sp properties we depart from the Dirac equation for
a free nucleon\footnote{Hereafter we set $G=c=\hbar=1$, where
$G$ is the gravitational constant.}, i.e.,
\begin{equation}
     (i\not \partial -m )\psi (x)=0,
\end{equation}
where $m$ is the free nucleon mass and $\psi (x)$
is  the nucleon field operator,
which is conventionally expanded in terms of plane wave states
and the Dirac spinors $u(p,s)$, and $ v(p,s)$, where
 $p=(p^0 ,{\bf p})$ is
a four momentum and  $s$ is the spin projection.
The relativistic energy is $E(p) =\sqrt{m^2 +|{\bf p}|^2}$.
To account for medium modifications to the free Dirac equation,
we introduce the notion of the self-energy $\Sigma (p)$, which
for nucleons can be written as
\begin{equation}
       \Sigma(p) =
       \Sigma_S(p) -\gamma_0 \Sigma^0(p)
       +\mbox{\boldmath $\gamma$}{\bf p}\Sigma^V(p).
\end{equation}
The momentum dependence of $\Sigma^0$
and $\Sigma_S$ is rather weak \cite{bm90,sw86}.
Moreover, $\Sigma^V << 1$, and it is customary to rewrite the self-energy
as
\begin{equation}
     \Sigma \approx \Sigma_S -\gamma_0 \Sigma^0 = U_S + U_V,
\end{equation}
where $U_S$ is an attractive
scalar field and $U_V$ is the time-like component
of a repulsive vector field.
The finite self-energy modifies the
free Dirac spinors as
\begin{equation}
   \tilde{u}(p,s)=\sqrt{\frac{\tilde{E}(p)+\tilde{m}}{2\tilde{m}}}
	  \left(\begin{array}{c} \chi_s\\ \\
	  \frac{\mbox{\boldmath $\sigma$}\cdot{\bf p}}
          {\tilde{E}(p)+\tilde{m}}\chi_s
	  \end{array}\right),
\end{equation}
where we let the terms with tilde
represent the medium modified quantities.
Here we have defined \cite{bm90,sw86}
\begin{equation}
   \tilde{m}=m+U_S,
\end{equation}
and
\begin{equation}
      \tilde{E}_{\alpha}=
      \tilde{E}(p_{\alpha})=\sqrt{\tilde{m}^2+{\bf p}_{\alpha}^2},
\end{equation}
and the sp potential is given by
\begin{equation}
   u_{\alpha} =\sum_{h\leq k_F} \frac{\tilde{m}^2}{\tilde{E}_h
	       \tilde{E}_{\alpha}}
	\bra{\alpha h}G(\omega =\tilde{\varepsilon}_{\alpha}
	+\tilde{\varepsilon}_h)\ket{\alpha h}_{\mathrm{AS}},
	\label{eq:urel}
\end{equation}
where $G$ is the relativistic $G$-matrix \cite{bm90}.
We can also express the sp potential in terms of the constants $U_S$ and
$U_V$,
\begin{equation}
   u_{\alpha} = \frac{\tilde{m}}{\tilde{E}_{\alpha}}U_S +U_V.
   \label{eq:sppotrel}
\end{equation}
The sp energies $\varepsilon$ can then be written as
\begin{equation}
   \tilde{\varepsilon}_{\alpha}=\tilde{E}_{\alpha} +U_V,
   \label{eq:sprelen}
\end{equation}
where we have used the continuous choice for the
single-particle energies \cite{ms89}.

Eqs.\ (\ref{eq:urel})-(\ref{eq:sprelen}) are solved self-consistently
starting
with adequate values for the scalar and vector components
$U_S$ and $U_V$. This iterative scheme is continued until these
parameters show little variation. The calculations are carried out in the
nuclear matter rest frame, avoiding thereby a cumbersome
transformation between the two-nucleon center-of-mass system
and the nuclear matter rest frame. The additional factors
$\tilde{m}/\tilde{E}$ in the above equations
arise due to the normalization of the
neutron matter spinors $\tilde{w}$, i.e.
$\tilde{w}^{\dagger}\tilde{w}=1$ \cite{bm90}. With the obtained sp energies,
we can calculate the relativistic energy per particle, see e.g.\ Refs.\
\cite{bm90,behoo94}. All relevant equations with a given proton-neutron
fraction are given in Ref.\ \cite{swk92}.

With these preliminaries, we present our results for the
energy per particle with various proton fractions
in Fig.\ \ref{fig:energy}.
For relatively small proton fractions,
the energy per
particle exhibits much the same curvature as the curve for pure neutron
matter at high densities, although the energy per particle is less repulsive
at high densities.
At lower densities, the situation is rather
different.
This is due to the contributions from various isospin $T=0$ partial waves,
especially the contribution from the $^3S_1$-$^3D_1$ channel, where the
tensor force component of the nucleon-nucleon potential provides
additional binding.
{}From Fig.\ \ref{fig:energy} we note that
with a proton fraction of $15\%$, the energy per particle
starts to become attractive at low densities (in the region $0.07$ fm$^{-3}$
to $0.3$ fm$^{-3}$). For larger proton fractions, additional
attraction to the energy
per particle is introduced.
The next step in our calculations is thus to evaluate the EOS and the total
mass
and radius of a neutron star from the above energies per particle with
the proton fractions shown in Fig.\ \ref{fig:energy}, in order
to see how different proton fractions influence the mass and radius of
neutron stars.
Here we
assume that the neutron stars we study exhibit an isotropic
mass distribution. Hence, from the general theory of relativity,
the structure of a neutron star is determined through the
Tolman-Oppenheimer-Volkov equation, i.e.\
\begin{equation}
   \frac{dP}{dr}=
    - \frac{\left\{\rho (r)+P(r) \right\}
    \left\{M(r)+4\pi r^3 P(r)\right\}}{r^2- 2rM(r)},
   \label{eq:tov}
\end{equation}
where $P(r)$ is the pressure and $M(r)$ is
the gravitional mass inside a radius $r$.
To obtain
observables like the mass and radius of a neutron star, we combine
Eq.\ (\ref{eq:tov}) with the equation of state (EOS), which is defined as
$P(n)=n^2 \left(\partial \epsilon/\partial n\right)$,
where $\epsilon ={\cal E}/A$ is
the energy per particle and $n$ is the particle
density.
Our EOS is valid in a limited density range from $0.1$ fm$^{-3}$
to $0.8$ fm$^{-3}$. It is therefore coupled to other equations
of state at higher and lower densities as outlined in Ref.\
\cite{behoo94}. Total masses and radii are calculated and parametrized
as functions of the central density $n_c$. These results are presented in
Figs.\ \ref{fig:fig2} and \ref{fig:fig3}, for the total mass and
radius, respectively.

The neutron star equation of state should probably be only
moderately stiff to support maximum neutron star masses of
only $1.9 M_{\odot}$ \cite{thorsett93}. From Figs.\ \ref{fig:fig2}
and \ref{fig:fig3} we see that our relativistic EOS for pure neutron
matter seems to be too stiff, since it gives a predicted
maximum mass of $M_{\mathrm{max}}\approx 2.4 M_{\odot}$ with
a corresponding radius of $R=12$ km. However, the EOS for neutron star
matter could be softened considerably due to pion or kaon
condensation because of the lower symmetry energy of nuclear matter,
and maximum masses are then reduced correspondingly
from the cases with no condensates. We see from Figs.\ \ref{fig:fig2}
and \ref{fig:fig3} that for our EOS the calculated maximum mass
can be reduced to $M_{\mathrm{max}}\approx 2.0 M_{\odot}$
with a corresponding radius of $R\approx 10$ km. Pions may be likely
to condense in neutron star matter because neutrons at the top of
the Fermi sea could decay to protons plus electrons \cite{mo88},
and kaon condensation is also believed to be a possible mechanism
which could be energetically favorable in the interior of neutron
stars \cite{brown93,thorson93,brown93b}. Both pion and kaon
condensation would then increase the proton abundance in the matter,
possibly up to more than $40\%$ protons, i.e.\ close to symmetric
nuclear matter, and produce a softer EOS and smaller maximum mass.
Our results show this, if we assume that e.g.\ kaon condensation is
a likely mechanism, an increased proton fraction results in smaller
masses. Our EOS, even with a proton fraction close to
symmetric nuclear matter, results in maximum masses which are slightly
above the experimental values \cite{thorsett93}. However, our calculation
of the EOS is to first order in the reaction matrix $G$, and we would
therefore expect that higher-order many-body contributions to
soften the EOS further. This was indeed shown in a preliminary study for
symmetric nuclear matter by Jiang {\em et al.} \cite{jiang93}. Although
only a set of higher-order contributions was considered, the above
authors obtain a softening of the relativistic EOS. The study of such
effects will be studied by us in future works.


\clearpage
\begin{figure}
      \setlength{\unitlength}{1mm}
      \begin{picture}(140,150)
      \end{picture}
\caption{The energy per particle for asymmetric nuclear matter
as function of the particle density for
different proton fractions.}
\label{fig:energy}
\end{figure}

\begin{figure}
      \setlength{\unitlength}{1mm}
      \begin{picture}(140,150)
      \end{picture}
\caption{$M/M_{\odot}$ for various proton fractions
as function of the central density $n_c$.}
\label{fig:fig2}
\end{figure}

\begin{figure}
      \setlength{\unitlength}{1mm}
      \begin{picture}(140,150)
      \end{picture}
\caption{The total radius  $R$ for different
proton fractions as function of the central density $n_c$.}
\label{fig:fig3}
\end{figure}

\end{document}